\documentclass[conference]{IEEEtran} 
\usepackage{amsmath}
\usepackage{algorithm}
\usepackage{amsmath,amssymb,amsfonts}
\usepackage{cite}
\usepackage{mathtools}
\usepackage[binary-units=true]{siunitx}
\DeclareSIUnit{\belmilliwatt}{Bm}
\usepackage{amsfonts}
\usepackage{amssymb}
\usepackage{graphicx}
\usepackage{dsfont}
\usepackage{subcaption}
\usepackage{comment}
\usepackage{bm}
\usepackage{mathtools}

\usepackage{textcomp}
\usepackage[export]{adjustbox}
\usepackage{float}
\usepackage{booktabs}
\usepackage{multicol}
\usepackage{xparse}
\usepackage[left=1.62cm,right=1.62cm,top=1.9cm]{geometry}
\usepackage{color,soul}
\usepackage[bottom]{footmisc}
\usepackage[binary-units=true]{siunitx}
\DeclareSIUnit{\belmilliwatt}{Bm}
\DeclareSIUnit{\dBm}{\deci\belmilliwatt}
\DeclareSIUnit[per-mode=symbol,per-symbol=p]{\Bps}{\byte\per\second}

\sethlcolor{yellow}
\usepackage{algpseudocode}

\usepackage{hyperref}

\makeatletter
\def\BState{\State\hskip-\ALG@thistlm}
\makeatother

\IEEEoverridecommandlockouts%\IEEEpubid{\makebox[\columnwidth]{ 978-1-6654-6483-3/23/\$31.00 ©2023 IEEE \hfill} \hspace{\columnsep}\makebox[\columnwidth]{ }}
\begin{document}

    \title{Proactive Radio Resource Allocation for 6G In-Factory Subnetworks}

\author{
\IEEEauthorblockN{Hossam Farag\IEEEauthorrefmark{1}\IEEEauthorrefmark{3}, Mohamed Ragab\IEEEauthorrefmark{2}\IEEEauthorrefmark{3}, Gilberto Berardinelli\IEEEauthorrefmark{1}, and \v{C}edomir Stefanovi\'{c}\IEEEauthorrefmark{1}}
\IEEEauthorrefmark{1} Department of Electronic Systems, Aalborg University, Denmark\\
\IEEEauthorrefmark{2} Propulsion and Space Research Center, Technology Innovation Institute, United Arab Emirates\\
\IEEEauthorrefmark{3} Department of Electrical Engineering, Aswan University, Egypt\\
Email: \{hmf, cs, gb\}@es.aau.dk, mohamedr002@e.ntu.edu.sg
}

	\maketitle
	
%%%%%%%%%%%%%%%%%%%%%%%%%%%%%%%%%%%%%%%%%%%%
%%%%%%%%%%%%%%%%%%%%%%%%%%%%%%%%%%%%%%%%%%%%
	\begin{abstract}
 6G In-Factory Subnetworks (InF-S) have recently been introduced as short-range, low-power radio cells installed in robots and production modules to support the strict requirements of modern control systems. Information freshness, characterized by the Age of Information (AoI), is crucial to guarantee the stability and accuracy of the control loop in these systems. However, achieving strict AoI performance poses significant challenges considering the limited resources and the high dynamic environment of InF-S. In this work, we introduce a proactive radio resource allocation approach to minimize the AoI violation probability. The proposed approach adopts a decentralized learning framework using Bayesian Ridge Regression (BRR) to predict the future AoI by actively learning the system dynamics. Based on the predicted AoI value, radio resources are proactively allocated to minimize the probability of AoI exceeding a predefined threshold, hence enhancing the reliability and accuracy of the control loop. The conducted simulation results prove the effectiveness of our proposed approach to improve the AoI performance where a reduction of 98\%  is achieved in the AoI violation probability compared to relevant baseline methods.

%  $\mathcal{O}_m$
	\end{abstract}
%%%%%%%%%%%%%%%%%%%%%%%%%%%%%%%%%%%%%%%%%%%%
%%%%%%%%%%%%%%%%%%%%%%%%%%%%%%%%%%%%%%%%%%%%
\section{Introduction}\label{sec:intro}
%\vspace{-3mm}
%%%%%%%%%%%%%%%%%%%%%%%%%%%%%%%%%%%%%%%%%%%%
Sixth-generation (6G) in-X subnetworks have recently been introduced as short-range, low-power radio cells designed to provide localized, high-performance wireless connectivity within entities such as industrial robots, vehicles, and the human body~\cite{sub1,sub2}. In a smart factory scenario, In-Factory Subnetworks (InF-S) are installed in robots and production modules encompassing a set of sensors and actuators locally connected to one or more access points that perform edge processing and control~\cite{InF}. This setup ensures that sensory data is transmitted and processed within sub-millisecond cycles, facilitating highly responsive control loops. In this context, the freshness of information becomes a key performance metric where real-time control decisions are based on the most recent and accurate data, minimizing the risk of errors caused by outdated information. While delay and packet delivery ratio are fundamental metrics for evaluating the performance of InF-S, neither metric accounts for the relevance or freshness of the delivered information. Even with low delay and high packet delivery ratio, a system may suffer from stale data if updates are not timely delivered, which could compromise the stability and efficiency of the control system. Age of Information (AoI)~\cite{AoI_def}, defined as the time elapsed since the generation of the last received update, serves as a key metric for quantifying information freshness. AoI incorporates all components of the latency from the moment of the data generation until its successful reception, offering a more holistic evaluation  compared to traditional metrics such as delay. While the average AoI reflects best-effort performance, ensuring reliability in terms of AoI (i.e., that AoI does not exceed a certain threshold) is crucial for effective and stable functioning of InF-S. However, guaranteeing reliable AoI performance represents a major challenge for InF-S considering their highly dynamic nature. In particular, mobility and  its corresponding fast changes in the interference level may limit the possibility for fulfilling the target AoI requirements. Therefore, the development of efficient and resilient algorithms that can adapt the allocation of radio resources to fluctuating network conditions, while adhering to stringent AoI requirements, is vital for these systems. 

While traditional resource allocation algorithms have been mainly based on hard-coded heuristics and optimization techniques, such as game theory and geometric programming methods which have shown effectiveness in specific scenarios~\cite{Heur}, the dynamic and intricate nature of InF-S requires more adaptive and data-driven approaches. Next generation wireless systems advocate for predictability by leveraging recent advances in Artificial Intelligence (AI) and Machine Learning (ML) to enable highly accurate predictions of channels, traffic and other key performance indicators~\cite{xurllc}. Unlike reactive approaches, the predictability feature enables proactive resource allocation to satisfy high levels of reliability guarantee even in dynamically changing network conditions. In the context of InF-S, if the AoI can be predicted with high accuracy, proactive resource allocation can be applied to minimize the likelihood of the future AoI exceeding a predefined threshold, thus achieving high reliability guarantee in terms of AoI. However, an accurate prediction of AoI necessitates comprehensive knowledge of network dynamics, including wireless channel conditions and interference levels, as well as past resource allocation decisions. Considering the highly dynamic nature of InF-S, it is challenging to acquire such knowledge a priori.
While Model Predictive Control (MPC) could be a viable solution  to control dynamic systems for optimized performance~\cite{MPC}, it requires an accurate mathematical model of the system which is unrealistic in industrial subnetworks. Recently, several solutions have been proposed for resource allocation targeting short-range low-power 6G in-X subnetworks~\cite{sub3, sub4, sub5, sub6, sub7}. The aforementioned works adopts ML-based algorithms, such as graph neural networks~\cite{sub3, sub4} and multi-agent reinforcement learning~\cite{sub5, sub6, sub7} for dynamic resource management. %for in-X subnetworks. 
While these studies have shown the potential for learning reasonably good solutions to radio resource allocation problems, they have not considered the AoI performance in the studied scenarios.
The AoI optimization problem in wireless networks has received a considerable attention~\cite{AoI1, hop4, opt1, opt2}. However, these works focus mainly on the average and peak AoI performance, which are insufficient metrics to characterize the real-time status updates in mission-critical applications as they do not account for extreme AoI events that occur with very low probabilities~\cite{schedule_4}. Moreover, these works have not bee designed for InF-S which have unique characteristics in terms of density and dynamic mobility.
In this work, we present a proactive AoI-aware resource allocation framework to address the critical challenges presented by InF-S to achieve a reliable AoI performance for modern control systems. The proposed approach adopts a decentralized, online learning strategy using Bayesian Ridge Regression (BRR)~\cite{BRR} to predict future AoI. %next time slot AoI.
The adopted BRR algorithm estimates the future AoI value by actively learning the complex and time-varying relationships between the subnetwork states (e.g., current AoI, transmission power, interference). Based on this predication, the subnetworks are proactively allocated radio resources to minimize the AoI violation probability. Specifically, the objective is to balance a trade-off between maximizing the knowledge gain about the system dynamics (exploration) and  improving the AoI reliability (exploitation), ensuring that the resource allocation decisions are both informed and adaptable to the system's non-stationary nature. The conducted simulations %results 
reveal the effectiveness of the proposed approach to improve the AoI reliability of InF-S compared to other baseline approaches

The rest of the paper is organized as follows. Section II presents the system model, followed by the proposed proactive resource allocation method in Section III. Performance evaluations are given in Section IV, and finally the paper is concluded in Section V.
%%%%%%%%%%%%%%%%%%%%%%%%%%
   \begin{figure}[t!] 
		\centering
		\includegraphics[width= 0.9\linewidth]{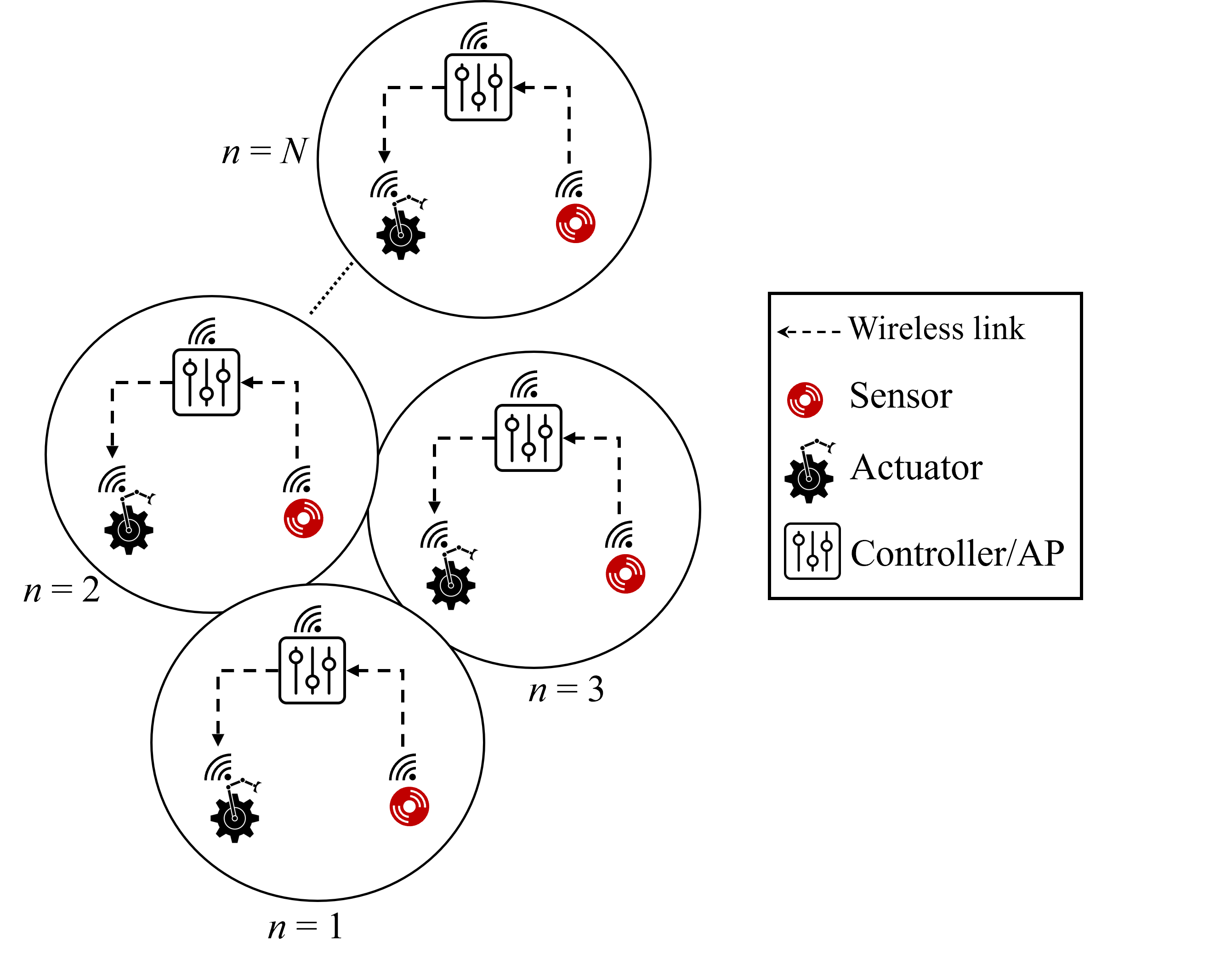}
		\caption{InF-S control system consisting of $N$ subnetworks.  \label{system}}
	\end{figure}
    \vspace{-8mm}
 %%%%%%%%%%%%%%%%%%%%%%%%%%

%%%%%%%%%%%%%%%%%%%%%%5%%%%
\section{System Model}
\label{system-model}
%\vspace{-2mm}
We consider an InF-S control system comprises a set $\mathcal{N}$ of $N$ independent mobile subnetworks as shown in Fig.~\ref{system}. Each subnetwork moves randomly with a speed  $v$ within the deployment area. The subnetworks provide wireless connectivity for a closed-loop control encompassing  sensors, actuators and a controller. The controller receives the measurements from the sensors (uplink) and generates control signals, based on the defined control objective, that are sent back to the actuators (downlink). Besides the conventional control tasks, the controller acts as an access point (AP) for the sensors and actuators in the subnetwork connecting it to a central controller. We assume the $N$ subnetworks to be time synchronized and share a set $\mathcal{B}$ of $B$ ($B<<N$) orthogonal resource blocks (RBs), each with an equal bandwidth $W$. The accuracy of control signals and the stability of the control loop are mainly influenced by the AoI of the received sensor data, hence, for the rest of the paper, we focus on the uplink transmissions. For the sake of simplicity, we assume that each subnetwork $n\in \mathcal{N}$ serves a single sensor node whose transmissions occupy the allocated RB. The time horizon is slotted into time slots of duration $\tau$ with $t$ representing the slot index. 
At each time slot $t$, the sensor of the subnetwork $n\in \mathcal{N}$ transmits at RB $b\in \mathcal{B}$ with one of the available $K+1$ power levels $P_n^b(t)=\{0, \frac{p}{K}, \frac{2p}{K}, ..., p\}$, where $p$ is the maximum transmission power. Hence, we can define the $B$-dimensional power allocation vector $\boldsymbol{P}_n(t)=[P_n^b(t)]_{b=1}^B\in \mathcal{P}$, where $\mathcal{P}$ is the set of all feasible power allocations. The sensor at each subnetwork $n$ samples the underlined process with a rate of $A$ packets per time slot, and the sampled packets are stored in  the sensor's output buffer whose dynamics are given as: $Q_n(t+1)=\mathrm{max}(Q(t)-R_n(t),0)+A$, where $Q(t)$ and  $R_n(t)$ are the buffer length  and  the transmission rate of the sensor $n$ at time slot $t$, respectively. The transmission rate $R_n(t)$ is given by
%%%%%%%%%%%%%%%%%%%%%%5%%%%%%%%%%%%%%%%%%%%%%%%%%5%%%%%%%%%%%%%%%%%%%%%%%%%%5%%%%
\begin{equation}\label{rate}
    R_n(t) = \frac{\tau}{L} \sum_{b \in \mathcal{B}} W \log_2 \left( 1 + \frac{P_n^b(t) h_{nn}^b(t)}{N_0 W + I_n^b(t)}\right),
\end{equation}
%%%%%%%%%%%%%%%%%%%%%%5%%%%%%%%%%%%%%%%%%%%%%%%%%5%%%%%%%%%%%%%%%%%%%%%%%%%%5%%%%
where $L$ is the packet size of the sensor data in bits, $h_{nn}^b(t)$ is the channel gain  of the sensor-controller link at subnetwork $n$, $N_0$ is the additive white Gaussian noise and $I_n^b(t)=\sum_{n'\neq n} P_{n'}^b(t) h_{n'n}^b(t)$ is the total interference experienced at the controller of subnetwork $n$ from other subnetworks over the RB $b$. Considering the small size of the subnetwork and the close proximity of sensors to their associated controllers, we adopt the Rician fading channel model~\cite{h_model} in this paper. We consider a single-shot transmission, i.e., no retransmission is considered for failed packet delivery. This assumption is based on the fact that transmitting a newly (fresh) arrived packet is more beneficial for the real-time performance of the controller than attempting to retransmit old/outdated packets.

The instantaneous AoI $\Delta_n(t)$ at the AP/controller of subnetwork $n$ at time slot $t$ can be defined as
%%%%%%%%%%%%%%%%%%%%%%5%%%%%%%%
\begin{equation}\label{aoi}
    \Delta_n(t)=\tau t-g_n(t),
\end{equation}
%%%%%%%%%%%%%%%%%%%%%%5%%%%%%%%
where $g_n(t)$ is the generation instant of the most recent update received by the controller of subnetwork $n$ at the beginning of time slot $t$. It can be noted that the index of the most recent received update mainly depends on the rate $R_n(t-1)$. Accordingly, and based on~\eqref{rate}, the AoI at each subnetwork is directly influenced by the channel gain, the interference and the allocated power. Therefore, we can model the future AoI $\Delta_n(t+1)$ (i.e., the value of the AoI in the next time slot) by a nonlinear function that captures the dynamics of the wireless link of subnetwork $n$ (AoI, resource allocation, power allocation vector, and channel unpredictability). Then,  we have
%%%%%%%%%%%%%%%%%%%%%%5%%%%%%%%
\begin{equation}\label{aoi}
    \Delta_n(t+1)= f_n(\Delta_n(t),\boldsymbol{P}_n(t)),  \,\, n=1, 2, ..., N,
\end{equation}
%%%%%%%%%%%%%%%%%%%%%%5%%%%%%%%
where $f_n(\Delta_n(t),\boldsymbol{P}_n(t))$ represents the nonlinear function that describes the dynamic system of subnetwork $n$, i.e., its instantaneous AoI $\Delta_n(t)$, channel gain, and interference. In other words, the nonlinear function $f_n$ maps $\boldsymbol{P}_n(t)$ (the control action) to $\Delta_n(t+1)$ (state of the system). Therefore, by learning the function $f_n$, according to \eqref{aoi}, the next slot AoI $\Delta_n(t+1)$ can be estimated. However, the function $f_n$ is unknown, since there is no prior knowledge about $h_{nn}^b(t)$, and $I_n^b(t)$\footnote {In this work, we do not consider adopting any CSI estimation techniques to avoid the corresponding complexity and extra overhead.}.

In this work, we adopt an online learning model to estimate and control $f_n$. The control part of $f_n$ implies the prediction of $\Delta_n(t+1)$ and accordingly allocate radio resources to satisfy a predefined AoI constraint. Specifically, our aim is to estimate $\Delta_n(t+1)$ based on historical data (learn the state of the dynamic system) and select power and RB (select the control action) that minimizes the AoI violation probability $ \Pr\left[ \hat{\Delta}_n(t+1) > \delta \right]$, where $\hat{\Delta}_n(t+1)$ is the estimated value of $\Delta_n(t+1)$ and $\delta$ is the predefined AoI threshold.
%%%%%%%%%%%%%%%%%%%%%%%%%%%%%%%%%%%%%%%%%
\section{AoI Prediction Using  Bayesian Ridge Regression}
\label{proposed}
%%%%%%%%%%%%%%%%%%%%%%%%%%%%%%%%%%%%%%%%%%
In this section, we formulate the problem of estimating $f_n$ as a regression problem, and adopt Bayesian Ridge Regression (BRR)~\cite{BRR} using the collected dataset which consists of $\Delta_n(t)$ and $\bm{P}_n(t)$. The selection of BRR is motivated by its compelling advantages in handling multicollinearity among features (e.g., power, interference, channel state), improving generalization through regularization, and crucially providing predictive uncertainty via a closed-form variance expression—essential for balancing reliability and exploration in the optimization objective. Denote $\bm{X}_t=[\Delta_n(t), \bm{P}_n(t)]$ as the input features, then BRR models the predicted AoI for the next time slot as
%%%%%%%%%%%%%%%%%%%%%%5%%%%%%%%
\begin{equation}\label{model1}
\hat{\Delta}_n(t+1) = w^T \Grave{\bm{X}_t} + b_0,
\end{equation}
%%%%%%%%%%%%%%%%%%%%%%5%%%%%%%%
where $w$ is the weight vector, $\Grave{\bm{X}_t}=\phi(\bm{X}_t)$ is the transformed feature vector, and $b_0$ is the bias term. In BRR, the weights $w$ are treated as random variables with a Gaussian prior, i.e., $w  \sim \mathcal{N}(0, \lambda^{-1} I)$, where $\lambda$ is a regularization parameter controlling the variance of the weights. The regularization parameter $\lambda$ helps in achieving a balance between bias and variance, improving generalization to unseen data. For a finite dataset $\mathcal{D}_n=\{\bm{x}_i, y_i\}_{i=1}^M$, where $\bm{x}_i=\Grave{\bm{X}_i}$, $y_i=\Delta_n(i+1)\}$, and $M$ is the dataset size, the posterior distribution of the weights $w$ is Gaussian:
%%%%%%%%%%%%%%%%%%%%%%5%%%%%%%%
\begin{equation}\label{weights}
w \mid D_n \sim \mathcal{N}(\hat{w}, \Sigma_w), 
\end{equation}
%%%%%%%%%%%%%%%%%%%%%%5%%%%%%%%
where the posterior mean $\hat{w}$ and covariance matrix $\Sigma_w$ are given by:
%%%%%%%%%%%%%%%%%%%%%%5%%%%%%%%
\begin{equation}\label{mean}
\hat{w} = \Sigma_w \bm{\Phi}^T \bm{y}
\end{equation}
%%%%%%%%%%%%%%%%%%%%%%5%%%%%%%%
%%%%%%%%%%%%%%%%%%%%%%5%%%%%%%%
\begin{equation}\label{variance}
\Sigma_w = (\lambda I + \bm{\Phi}^T \bm{\Phi})^{-1},
\end{equation}
%%%%%%%%%%%%%%%%%%%%%%5%%%%%%%%
with $\bm{y}=[\Delta_n(1), \Delta_n(2), ..., \Delta_n(M)]$ is the vector of observed AoI values. $\bm{\Phi}$ denotes the transformed design matrix where each row is $\Grave{\bm{X}_t}$, which corresponds to the transformed feature vector for the $t$-th observation.  The transformation matrix $\bm{\Phi}$  maps the original feature vector $\bm{X}_t$ into a higher-dimensional feature space where linear relationships might better approximate the underlying non-linear dynamics. The transformation matrix $\bm{\Phi}$ is obtained by applying a transformation function $\phi(.)$ (e.g., polynomial features, kernel functions~\cite{BRR}). So, $\bm{\Phi}$ is given as
\begin{equation}
\bm{\Phi} = 
\begin{bmatrix}
\phi(X_1)^T \\
\phi(X_2)^T \\
\vdots \\
\phi(X_M)^T
\end{bmatrix}
=
\begin{bmatrix}
\phi_1(X_1) & \phi_2(X_1) & \cdots & \phi_r(X_1) \\
\phi_1(X_2) & \phi_2(X_2) & \cdots & \phi_r(X_2) \\
\vdots      & \vdots      & \ddots & \vdots      \\
\phi_1(X_M) & \phi_2(X_M) & \cdots & \phi_r(X_M)
\end{bmatrix},
\end{equation}
where $r$ is the number of features in the transformed space. BRR will provide a predictive distribution of $\hat{\Delta}_n(t+1)$, that is, for a general input $\bm{x_*}$ and a given $\mathcal{D}_n$, BRR will provide a statistical description of the corresponding output $y_*$ (i.e., $\mathrm{Pr}(y_*|\bm{x_*}, \mathcal{D}_n$). Using the standard tools of Bayesian statistics, the predicted AoI value $\hat{\Delta}_n(t+1)$ has a normal distribution which is given by
%%%%%%%%%%%%%%%%%%%%%%5%%%%%%%%%%%%%%%%%%%%%%%%%%%%%%5%%%%%%%%%%%%%%%%%%%%%%%%%%%%%%5%%%%%%%%
\begin{equation} 
    \hat{\Delta}_n(t+1) \mid X_t, D_n \sim \mathcal{N}\left(\mu_{\hat{\Delta}_n(t+1)}, \sigma^2_{\hat{\Delta}_n(t+1)}\right),
\end{equation}
%%%%%%%%%%%%%%%%%%%%%%5%%%%%%%%%%%%%%%%%%%%%%%%%%%%%%5%%%%%%%%%%%%%%%%%%%%%%%%%%%%%%5%%%%%%%%
where the predictive mean $\mu_{\hat{\Delta}_n(t+1)}$ and the predictive variance $\sigma^2_{\hat{\Delta}_n(t+1)}$ are given by 
%%%%%%%%%%%%%%%%%%%%%%5%%%%%%%%
\begin{equation} \label{avg}
   \mu_{\hat{\Delta}_n(t+1)} = \hat{w}^T \Grave{\bm{X}_t} + b_0,
\end{equation}
%%%%%%%%%%%%%%%%%%%%%%5%%%%%%%%
%%%%%%%%%%%%%%%%%%%%%%5%%%%%%%%
\begin{equation} \label{STD}
\sigma^2_{\hat{\Delta}_n(t+1)} = \sigma^2 + \Grave{\bm{X}_t}^T \Sigma_w \Grave{\bm{X}_t},
\end{equation}
%%%%%%%%%%%%%%%%%%%%%%5%%%%%%%%
where $\sigma^2$ is the noise variance in the observations. 

As mentioned earlier, the estimated value of the future AoI $\hat{\Delta}_n(t+1)$ could be used to proactively allocate radio resources in order to minimize the AoI violation probability at subnetwork $n$, i.e., $\Pr\left[\hat{\Delta}_n(t+1) > \delta\right]$. However, accurate estimation of $\hat{\Delta}_n(t+1)$ requires precise knowledge of the system dynamics, i.e., $f_n$ in \eqref{aoi}. Moreover, an action (i.e., resource allocation), that enhances the knowledge about the system dynamics may not necessarily be the optimal one to minimize the AoI violation probability. Therefore, the aim is to balance the tradeoff between improving the AoI performance and acquired knowledge about the system dynamics. Hence, the %AoI-driven 
optimization problem can be defined as follows 
%%%%%%%%%%%%%%%%%%%%%%5%%%%%%%%
\begin{equation}\label{aoi-opt}
\begin{split}
&\min_{\bm{P_n(t)} \in \mathcal{P}}  \Pr\left[ \hat{\Delta}_n(t+1) > \delta\right] \\
&\mathrm{subject\,\,to\,\,} \eqref{aoi}.
\end{split}
\end{equation}
%%%%%%%%%%%%%%%%%%%%%%5%%%%%%%%
The problem in \eqref{aoi-opt} represents the objective of minimizing the AoI violation probability where the system dynamics constituting the constraint that should be learned  by selecting the appropriate action (RB and transmission power). The learning objective can be given by
%%%%%%%%%%%%%%%%%%%%%%5%%%%%%%%
\begin{equation}\label{info-opt}
\max_{\bm{P_n(t)} \in \mathcal{P}} \mathcal{I}_n (t) \left(\hat{\Delta}_n(t+1);\Delta_n(t), \bm{P_n(t)}, \mathcal{D}_n \right),
\end{equation}
%%%%%%%%%%%%%%%%%%%%%%5%%%%%%%%
where $\mathcal{I}_n\geq 0$ denotes the amount of expected knowledge gained at time $t$ for a given $\mathcal{D}_n$, $\Delta_n(t)$ and $\bm{P}_n(t)$. In order to quantify the knowledge gain $\mathcal{I}_n$ in \eqref{info-opt}, different tools from the field of information theory could be applied, such as entropy, mutual information, relative entropy, and Fisher information~\cite{linear}. In this paper, we consider entropy to quantify $\mathcal{I}_n(t)$. It can be noted that, based on entropy power from the observations $\{\Delta_n(t+1), \bm{P_n(t)}\}$, a lower bound of $\mathcal{I}_n(t)$ is directly proportional to the variance function $\sigma^2_{\hat{\Delta}_n(t+1)}$, which reflects the uncertainty of the prediction value provided by $f_n$. Hence, the objective in \eqref{info-opt} is equivalent to maximizing $\sigma^2_{\hat{\Delta}_n(t+1)}$. Accordingly, we can formulate the following dual-objective function
%%%%%%%%%%%%%%%%%%%%%%5%%%%%%%%
\begin{equation}\label{dual-opt}
\min_{P_k(t) \in \mathcal{P}} \ \alpha_c \Pr\left[\hat{\Delta}_n(t+1) > \delta\right] - \alpha_i \sigma^2_{\hat{\Delta}_n(t+1)}.
\end{equation}
%%%%%%%%%%%%%%%%%%%%%%%%%%%%%%%
The formulated optimization problem in \eqref{dual-opt} describes the tradeoff between improving AoI (exploitation) and enhancing the knowledge about the system dynamics (exploration), where $\alpha_c$ and $\alpha_i$ are non-negative weighting factors that capture the balance of the target exploitation-exploration tradeoff.

In this work, we advocate for a distributed learning architecture rather than a centralized one to mitigate the incurred communication overhead of exchanging the AoI values and resource allocation decisions at each time slot.  Algorithm~1 presents our adopted decentralized learning and proactive resource allocation approach to solve the optimization problem in ~(\ref{dual-opt}).
%%%%%%%%%%%%%%%%%%%%%%%%%%%%%%%

\begin{algorithm}[t!]
\caption{Distributed BRR-based active learning for InF-S}
\begin{algorithmic}[1]
\State \textbf{Inputs:} Set of possible power actions $\mathcal{P}$, parameters $\alpha_c$, $\alpha_i$, and dataset size $M$.
\State \textbf{Initialization:} Set $\Delta_n(0) \gets 0$, initialize $\mathcal{D}_n \gets \emptyset$, and randomly select $\bm{P}_n(0) \in \mathcal{P}$.
\For{each time slot $t = 1, 2, \dots$}
    \State Record the current AoI $\Delta_n(t)$.
    \State Update the dataset: $\mathcal{D}_n \gets \mathcal{D}_n \cup \{(\Delta_n(t-1), P_n(t-1)), \Delta_n(t)\}$.
    \If{$|\mathcal{D}_n| > M$}
        \State Remove the oldest entry from $\mathcal{D}_n$.
    \EndIf
    \State Compute $\mu_{\hat{\Delta}_n(t+1)}$ \eqref{avg} and $\sigma^2_{\hat{\Delta}_n(t+1)} \eqref{STD}\,\,\forall \bm{P}_n(t) \in \mathcal{P}$.
    \State Evaluate $\Pr [\hat{\Delta}_n(t+1) > \delta]\,\,\forall \bm{P}_n(t) \in \mathcal{P}$.
    \State Select the action $\bm{P}^*_n(t)$ that minimizes the objective function in~\eqref{dual-opt}.
\EndFor
\end{algorithmic}
\end{algorithm}
%%%%%%%%%%%%%%%%%%%%%%%%%%%%%%%
%%%%%%%%%%%%%%%%%%
\begin{table}[t!]
		\centering
		\caption{Simulation parameters}
		\label{t1}
		\begin{tabular}{ll}
			\toprule
			Parameter & Value \\
				\midrule
			Deployment area & $20$m $\times$ $20$m\\
            Number of Subnetworks ($N$) & $20$\\
            Subnetwork radius ($R_{sub}$) & $2$m\\
            Sensor to AP minimum distance ($d_{min}$) & $1$m\\
            Velocity of a subnetwork $v$ & $2$ m/s\\
            Number of RBs ($B$)& $5$\\
            Packet size ($L$)& $100$ bytes\\
			Available power levels ($K+1$) & 3 \\
			  Maximum transmission per RB ($p$) & \num {10} dBm \\
            AWGN power spectral density ($N_0$) & \num {-174} dBm/Hz \\  
			Time slot duration ($\tau$) & \num {3} ms\\
            AoI threshold ($\delta$) & \num {10} ms\\
			\bottomrule
		\end{tabular}	
	\end{table}
%%%%%%%%%%%%%%%%%%%%%%%%
%%%%%%%%%%%%%%%%%%%%%%%%%%%%%%%
\section{Performance Evaluation}
\label{results}
%%%%%%%%%%%%%%%%%%%%%%%%%%%%%%%

In this section, we evaluate the performance of the proposed method via comprehensive Monte Carlo simulations in MATLAB using the parameters listed in Table~\ref{t1}. 
%%%%%%%%%%%%%%%%%%%%%%%%%%%%%%%
\subsection{Simulation Setup}
%%%%%%%%%%%%%%%%%%%%%%%%%%%%%%%
We consider InF-S control system consisting of $20$ mobile subnetworks randomly distributed in 20m $\times$ 20m deployment area, which is a typical scenario in industrial subnetworks~\cite{InF, par1}. At each subnetwork, the AP/controller is positioned at the center of the subnetwork, which is represented by a circular coverage area with radius $R_{sub}$, and the sensor is located at a distance $d$ from the AP, ensuring a minimum proximity of $d_{min}$.  We consider a restricted random direction model (RDM)~\cite{mobility}. At the beginning of each simulation run, the subnetworks are randomly distributed in the deployment area, then they move with a constant speed $v$ in random directions. A subnetwork randomly changes its direction when it reaches a boundary or if its distance to any other subnetwork is less than 1.5 m. We adopt the channel model for indoor factory scenarios defined in~\cite{3GPPURLLC}. The model considers Rician fading with a K-factor of 7 representing a high likelihood of line-of-sight conditions in the subnetworks. In addition, we consider the path loss model defined in~\cite{3GPPURLLC}, incorporating a shadow fading model with a standard deviation of 7~dB.

We compare the proposed approach with two baselines. The first one is based on a random resource allocation (referred as Default in the results) where the power and RBs are allocated randomly among the subnetworks without applying BRR. The second one is a greedy version of the proposed method (referred as greedy in the results) where the exploration phase is excluded during the learning phase (i.e., $\alpha_i=0$).
%%%%%%%%%%%%%%%%%%%%%%%%%%%%%%%
\subsection{Results and Discussion}
%%%%%%%%%%%%%%%%%%%%%%%%%%%%%%
 %%%%%%%%%%%%%%%%%%%%%%%%%%
   \begin{figure}[t!] 
		\centering
		\includegraphics[width= 0.9\linewidth]{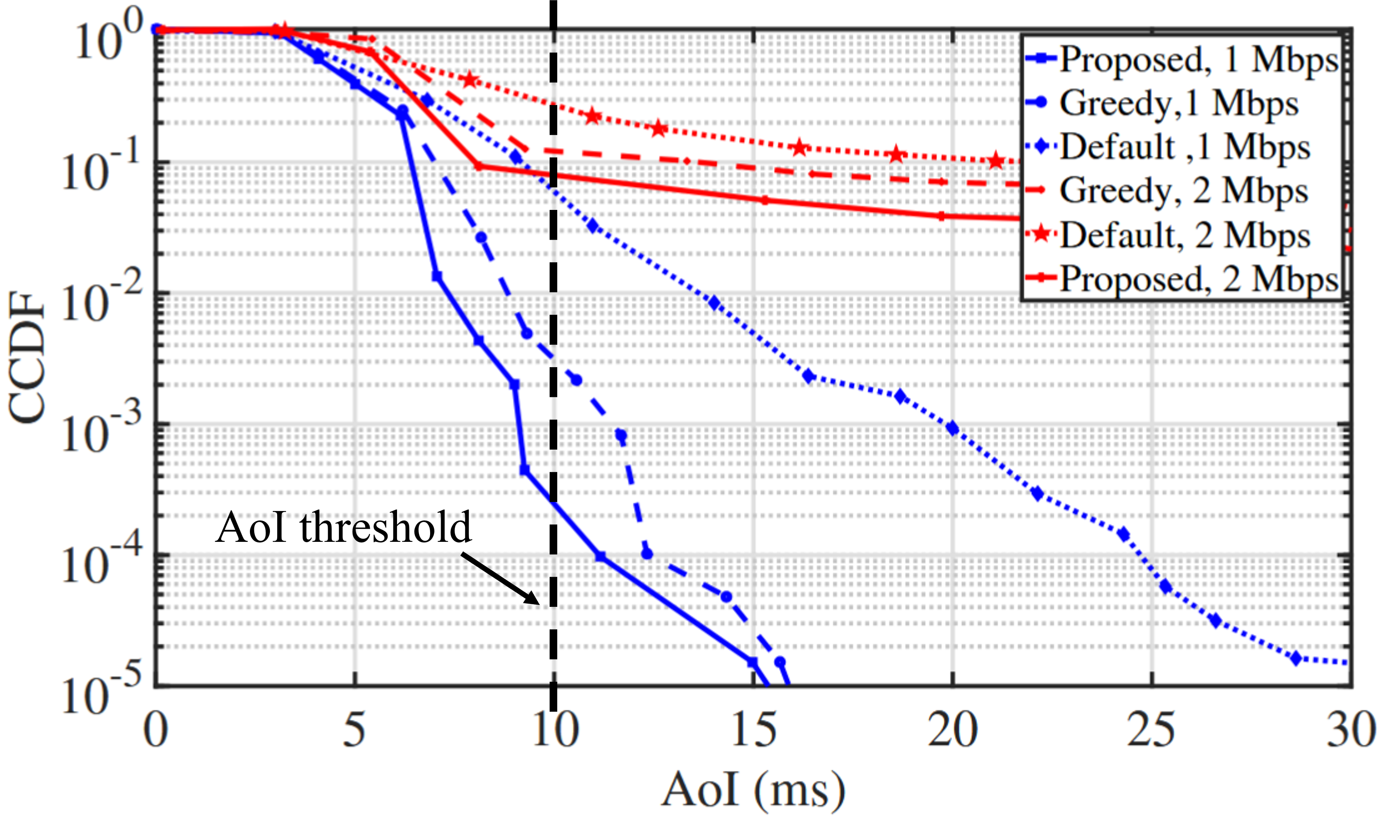}
		\caption{CCDF comparison of the AoI with $M=300$ and $\delta=10$ ms.  \label{CCDF}}
        \vspace{-5mm}
	\end{figure}
 %%%%%%%%%%%%%%%%%%%%%%%%%%

 Fig.~\ref{CCDF} compares the Complementary Cumulative Distribution Function (CCDF) of the AoI of the different resource allocation strategies under different sampling rates  of 1 Mbps and 2 Mbps. Compared to the default resource allocation strategy, the results clearly demonstrate the effectiveness of the adopted dynamic learning strategy (proposed and greedy methods) to improve the AoI performance of mobile InF-S under different sampling rates. For instance, the proposed method improves the AoI violation probability of the default method by almost $98\%$ with 1 Mbps sampling rate. The figure also shows that the sampling rate has a direct effect on the AoI violation probability mainly due to the excess queuing time where a packet spends more time before its transmitted to the controller, which could likely lead to violating the AoI threshold.
  %%%%%%%%%%%%%%%%%%%%%%%%%%
   \begin{figure}[t!] 
		\centering
		\includegraphics[width= 1\linewidth]{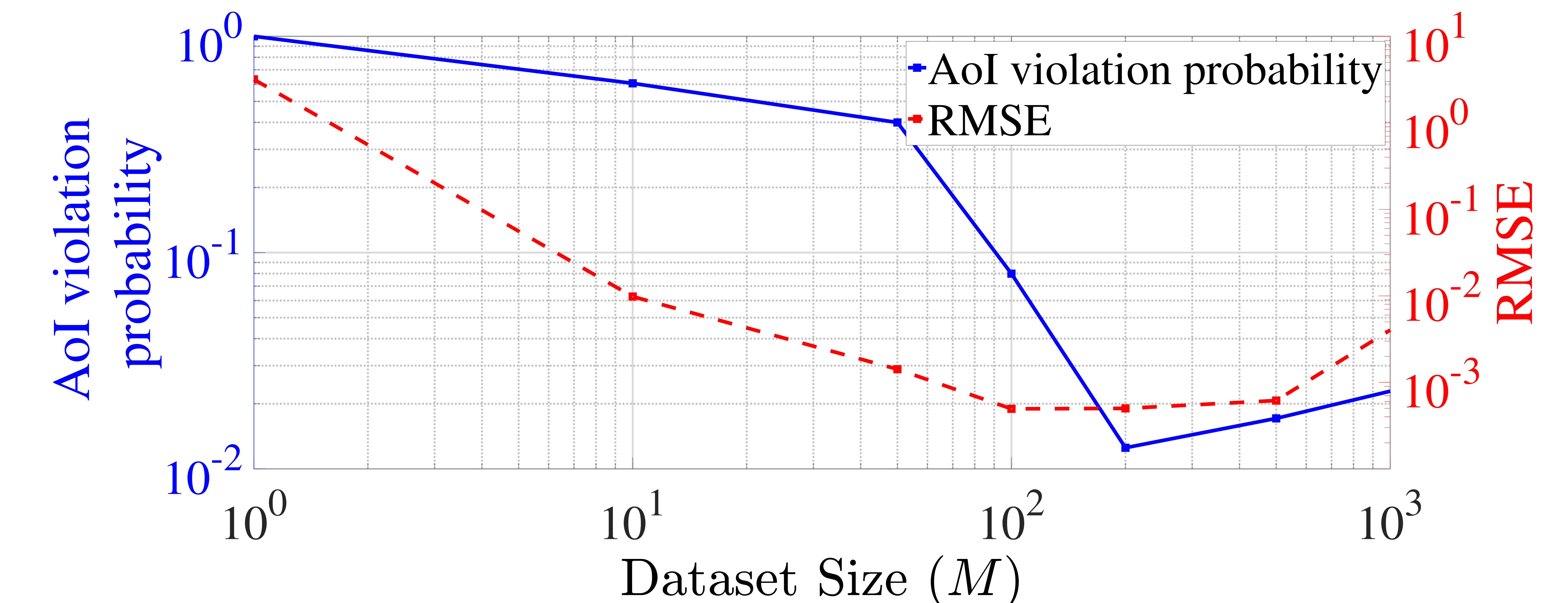}
		\caption{RMSE and AoI violation probability under varying dataset sizes.  \label{rmse}}
	\end{figure}
 %%%%%%%%%%%%%%%%%%%%%%%%%%

Fig.~\ref{rmse} evaluates the effect of the size of the dataset $M$ on the learning and AoI performance of the proposed method. The learning performance is measured in terms of the prediction accuracy, which is quantified using the Root Mean Square Error (RMSE):
%%%%%%%%%%%%%%%%%%%%%%5%%%%%%%%
\begin{equation}\label{accuracy}
\mathrm{RMSE}=\sqrt{\sum_t\left(\mu_{\hat{\Delta}_n(t+1)}-\Delta_n(t+1)\right)^2/T}.
\end{equation}
%%%%%%%%%%%%%%%%%%%%%%5%%%%%%%%
It can be noted that, while increasing the dataset size improves the prediction performance (RMSE) with $M$ up to 200, beyond this value, the RMSE values starts to increase along with a notable degradation on the AoI violation probability. This is mainly due to the fact that, for higher values of $M$, due to the high dynamic nature of the InF-S, the dataset my include irrelevant and outdated samples from the learning environment that mislead the BRR model with uncorrelated observations, hence producing inaccurate predictions. Another positive side that can be revealed from this observation is that the computational complexity of the BRR algorithm can be relaxed as there is no need for a large size of dataset to achieve accurate estimation of future AoI values, which somwhat relxes the computational complexity of the adopted BRR algorithm.
   %%%%%%%%%%%%%%%%%%%%%%%%%%
   \begin{figure}[t!] 
		\centering
		\includegraphics[width= 1\linewidth]{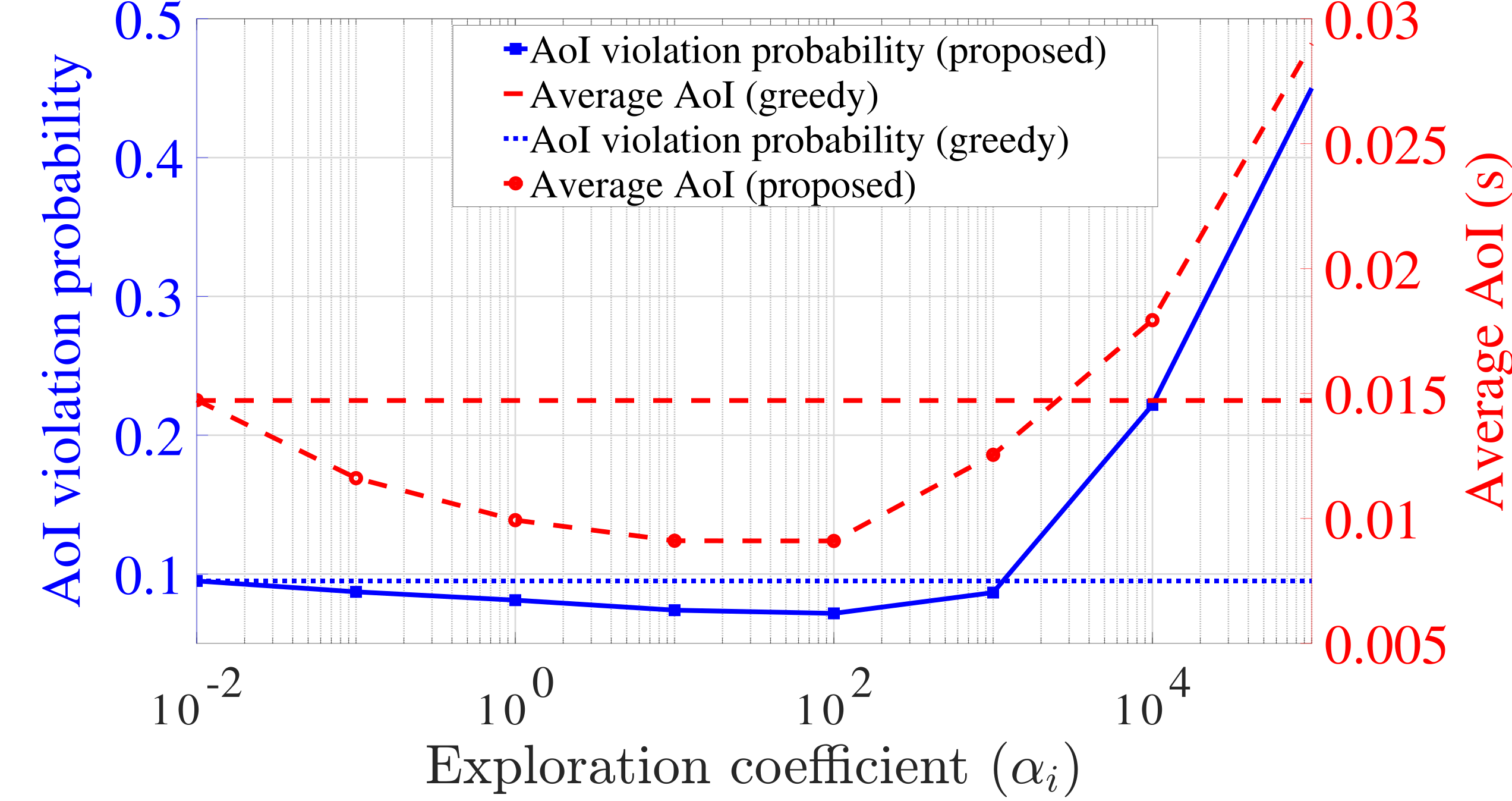}
		\caption{Average AoI and AoI violation probability under varying $\alpha_i$.  \label{explore}}
	\end{figure}
 %%%%%%%%%%%%%%%%%%%%%%%%%%

In Fig.~\ref{explore}, we evaluate the effect of the exploration parameter $\alpha_i$ on the AoI performance including the greedy case with no exploration. When the BRR favors more the exploration behavior (i.e., increasing $\alpha_i$), the BRR algorithm tends to select the actions that enhance the knowledge about the dynamic system, which in turns improves the prediction accuracy and accordingly decreases the average AoI and the AoI violation probability as dsepcited by Fig.~\ref{explore}. However, prioritizing exploration over exploitation beyond a certain level ($\alpha_i>100$) can significantly degrade the AoI performance, since the BRR will unlikely to select the control actions that minimize the AoI violation probability. The results indicates that the learning metrics $\alpha_i$ and $\alpha_c$ should be carefully selected to achieve a balance regarding the exploration-exploitation tradeoff.
 
%%%%%%%%%%%%%%%%%%%%%%%%%%%%%%%
\section{Conclusion and Future Work}
\label{sec:conclusions}
%%%%%%%%%%%%%%%%%%%%%%%%%%%
This work introduced a proactive resource allocation approach to improve the AoI performance of mobile In-Factory Subnetworks. The proposed solution is based on an online learning approach levering the BRR algorithm with the objective to minimize the AoI violation probability while enhancing the model knowledge about the system dynamics. By predicting the next-slot AoI, radio resources are proactively allocated to fulfill the predefined AoI constraint. The obtained results showed that the proposed method can outperform relevant baseline methods in terms of the AoI violation probability while introducing acceptable level of complexity. The results also revealed that more investigations are needed to address how to optimally adjust the exploitation and exploration weights for optimal AoI performance. Furthermore, evaluating the AoI performance of the proposed method by considering both inter- and intra-subnetwork interference is another potential topic for future work.

%%%%%%%%%%%%%%%%%%%%%%%%%%%%
%\section*{Acknowledgement}
%%%%%%%%%%%%%%%%%%%%%%%%%%%
%This paper has received funding from the European Union’s Horizon 2020 research and innovation programme under Grant Agreement number 856967. 

	\bibliographystyle{IEEEtran}
\bibliography{mybib}
	
\end{document}